# On a universal relation for defects in solids.

E.S. Skordas

*Section of Solid State Physics and Solid Earth Physics Institute, Department of Physics, National and Kapodistrian University of Athens, Panepistimiopolis, Zografos 157 84, Athens, Greece.*

**Abstract**

We show that the defect data parameters related to various defect processes, e.g., formation, migration, dielectric relaxation parameters, obey a universal law. In particular, the defect entropies scale with the defect enthalpies irrespective of the process considered. A concrete example is given here for $SrF_2$ by considering the dielectric relaxation parameters ($R_1$ relaxation mechanism) for crystals doped with trivalent ions of Ce, Eu and Gd, parameters for the anion Frenkel formation as well as for the migration of anion vacancy and the anion interstitial motion.

----------------------------------

eskordas@phys.uoa.gr

# 1. Introduction

Halides in general have variety of important roles in material science. First, they serve as test materials for condensed matter theoretical predictions of ionic materials. Second, since they are more compressible than oxides and silicates, they exhibit larger strains upon the application of a relatively lower pressure range. Thus, they have been long used as mineralogical models for the behavior of ionically bonded minerals in the Earth's mantle and as test materials for the discovery of high-pressure equations of state [1]. Third, since they are usually elastically compressible, the halides are frequently used for the study of high-pressure and high-temperature materials behavior [2, 3].

When focusing on the fluorite-structured halides, there are additional important reasons for intensifying their study during the last three decades. Chief among these is the discovery that these systems exhibit superionic conductivity at high temperatures [4-7] suitable for solid state battery sensors [8]. For example, we mention the case of $PbF_2$ which has attracted a major interest [9-15], because beyond the fact that its $\beta$-phase exhibits high superionic conductivity with low superionic transition temperature [6, 7], its $\alpha$-phase is a strong scintillator in detector technology [16]. The latter is so since $\alpha$-$PbF_2$ exhibits short radiation lengths, short decay time, high light yield and a good radiation hardness [17, 18], which are prerequisites for scintillating detectors to be used in high energy physics experiments in which the particle energy exceeds GeV [19]. Furthermore, electrical relaxation measurements (see also below) at high pressures on $PbF_2$ doped with La and Ce showed [20] a single, strong relaxation peak the relaxation time of which decreases with increasing pressure. Such a behavior, which is equivalently interpreted as being due to a negative activation volume for the defect motion (see also Refs. [21,22]), is of major importance, because it provides the basis [23] for the generation of the electric signals observed before earthquakes [24, 25]. These signals are emitted when the stress in the focal area –when it lies inside the Earth's crust– approaches a critical point (second order phase transition [26]).

Characteristic example of fluoride-structure halides is the case of alkaline-earth fluorides, which beyond the aforementioned superionic behavior and other applications of practical interest (e.g., $CaF_2$ is commonly designated as a lens material for photolithography at wavelengths in the deep ultraviolet region, see Ref. [27] and references therein) are important in the following respect: Understanding the pressure-induced phase transition of these materials will be helpful for understanding of phase transformations in the deep Earth. Therefore, the interest on the electronic and optical properties of these halides [28] as well as on their structural stability [29] still remains high. Of particular importance is the study of their defects on which the present paper is focused on. Various experimentally techniques showed [8] that the predominant intrinsic point defects are anion Frenkel pairs and that ionic transport takes place through migration of (free) anion vacancies and (free) interstitials. The formation (*f*) and migration (*m*) parameters of these defects will be designated with a superscript *f* and *m*, respectively. For example, $h^f$ and $s^f$ stand for the formation enthalpy and the formation entropy of a Frenkel pair.

Beyond the intrinsic point defects, however, extrinsic defects may be formed when aliovalent impurities are present. The activation (*act*) of these defects is associated with parameters, i.e., the enthalpy $h_i^{act}$ and the entropy $s_i^{act}$, which (not only are different from those of the intrinsic defects but also) depend in general on the aliovalent impurity which is present in each case. Thus, when we have different aliovalent impurities that produce extrinsic defects activated in the same lattice (matrix), the question raises whether the corresponding parameters (i.e., enthalpies and entropies) obey a general relation. Here, we shall show that such a relation does exist. This will be described in the next Section. In Section 3, we shall consider as a typical example the case of SrF$_2$.

**2. The universal relation.**

Before proceeding, we clarify that the parameters $s^{act}$ and $h^{act}$ may depend in general on temperatures, but it has been demonstrated [23] that they must obey the following rule: They may both increase or decrease upon increasing the temperature $T$.

According to a thermodynamical model, termed c$B\Omega$ model, the defect Gibbs energy $g^i$ is interconnected with the bulk properties of the solid through the relation [23]

$$g^i = c^i B\Omega \tag{1}$$

where $B$ stands for the isothermal bulk modulus, $\Omega$ the mean volume per atom, $c^i$ is dimensionless independent of temperature and pressure, and the superscript $i$ refers to the defect process under consideration, i.e., $i= f, m, act$, for the formation, migration, and activation, respectively, as mentioned. Since [23] $s^i = -\left(dg^i/dT\right)_p$ (where $p$ denotes the pressure), we differentiate Eq. (1) with respect to temperature, thus obtaining an expression for $s^i$ and then, using the relation $h^i = g^i + Ts^i$, we get the relevant expression for the defect enthalpy $h^i$. These two expressions for $s^i$ and $h^i$ lead to [23]

$$\frac{s^i}{h^i} = -\frac{\beta B + \left.\frac{dB}{dT}\right|_p}{B - T\beta B - T\left.\frac{dB}{dT}\right|_p}, \qquad (2)$$

where $\beta$ is the thermal volume expansion coefficient. The validity of Eq (2) has been checked for various defect processes in a variety of solids [30-34], e.g., ionic crystals, metals, noble gas solids as well as mixed alkali halides (where the $\beta, B$ values can be estimated from the corresponding values of the end members [23, 35]).

Let us label $F$ the right hand side of Eq (2), which solely depends on quantities of the matrix material:

$$F \equiv -\frac{\beta B + \left.\frac{dB}{dT}\right|_p}{B - T\beta B - T\left.\frac{dB}{dT}\right|_p}, \qquad (3)$$

Hence, Eqs (2) and (3) reveal that when plotting –for various defect processes– the experimental value of $h^i$ versus $s^i/F$, we should find a straight line with a slope equal to unity. This should hold for all solids, thus revealing universality.

### 3. The data analyzed. Results

As a typical example, we shall indicate here, as mentioned, how the aforementioned universality appears in the case of SrF$_2$. In particular we consider the following two data sets:

The first data set refers to dielectric relaxation measurements of SrF2 doped with rare earth ions. For each trivalent dopant ion, and for reasons of charge compensation, a fluorine interstitial ($F^-$) is produced, thus electric dipoles are formed. The (re)orientation of these dipoles has been studied by measuring the real ($\varepsilon'$) and imaginary ($\varepsilon''$) parts of the dielectric constant [36, 37]. These measurements lead to the determination of the activation energy $E$ and the preexponential factor $\tau_0$ of the Arrhenius relation [38, 39]:

$$\tau = \tau_0 \exp(E/kT), \tag{4}$$

where $\tau$ stands for the relaxation time and $k$ the usual Boltzmann constant. In the usual case of a linear $\ln \tau$ versus $1/T$ plot, the energy $E$ is the enthalpy $h^m$ for the corresponding migration (re-orientation) process. Here, we shall focus on the so called $R_1$-relaxation that is associated with complexes (electric dipoles) in which the $F^-$ occupies a nearest neighbor (*nn*) interstitial position with respect to trivalent earth ion. The $\tau_0$ and $E$ –values reported in Refs [40, 41] for the trivalent ions $Ce^{3+}$, $Eu^{3+}$ and $Gd^{3+}$ are inserted in Table 1. The procedure to extract the migration entropy $s^m$ from the experimental data goes as follows (e.g., see Chapter 11 of Ref [23]): The relaxation time $\tau$ can be generally written as

$$\tau = (\lambda \nu)^{-1} \exp(g^m/kT), \tag{5}$$

where $\lambda$ denotes the number of jump paths accessible to the jumping species with an attempt frequency $\nu$ and $g^m$ the corresponding migration (*m*) Gibbs energy. Since the latter can be written as $g^m = h^m - Ts^m$, and the migration entropy $s^m$ is given by $s^m = -(dg^m/dT)_p$, Eq (5) leads to

$$\tau = (\lambda \nu)^{-1} \exp(-s^m/k) \exp(h^m/kT) \tag{6}$$

By comparing Eqs. (1) and (6), we conclude that in the case of a linear $\ln \tau$ versus $1/T$ plot, the activation energy $E$ is just the migration enthalpy $h^m$, and the pre-exponential factor $\tau_0$ is interconnected to $s^m$ through

$$s^m = k \ln\left[\tau_0^{-1}/(\lambda \nu)\right] \tag{7}$$

For the R$_1$ relaxation we have [23] the value $\lambda = 2$. As for the quantity $v$ we use the approximation $v = v_{T_0}$ ($k \to 0$) where $v_{T_0}$ denotes the frequency of the long wavelength transverse optical mode. For the latter quantity Lowndes [42] measured the values 397 cm$^{-1}$ and 395 cm$^{-1}$ for temperatures T=5K and 290K, respectively. (These are comparable with the values measured in Refs [43] and [44]). Taking their mean value 396 cm$^{-1}$, Eq (7) leads to the experimental $s^m$ values inserted in Table 1. In the same Table, we also give the $F$-value when using the expansivity and the compressibility ($=1/B$) data quoted in Ref. [45].

The second data set used here refers to the defect entropies and enthalpies, as compiled by Bollmann [46], that have been deduced from the analysis of the conductivity plots $\ln(\sigma T)$ vs $1/T$ in a way described in Ref. [39]. They are also inserted in Table 1 and correspond to the following processes: anion-Frenkel formation, anion vacancy migration and anion interstitial motion.

Using the aforementioned values coming from both the dielectric relaxation and the conductivity measurements, we plot in Fig. 1, the $s^i/F$-values versus $h^i$. In the same plot, for the sake of reader's convenience, we also plot a straight line with slope equal to unity. An inspection of this figure reveals a satisfactory agreement with the behaviour expected for reasons explained in the previous Section. The same holds for other solids as well, i.e., when considering noble gas solids the ordinates of the points ($s^i/F$, $h^i$) are smaller than those in Fig. 1 by almost one order of magnitude, but these points should lie on the *same* straight line of unity slope (while for oxides of high melting temperatures the ordinates are one order of magnitude larger, but still exhibit the same behaviour as that in Fig. 1).

## 4. Conclusion

Here, we showed that for different defect processes in the same material the ratio $s^i/h^i$ should be equal to a bulk quantity $F$. This finally reflects that when plotting $s^i/F$ versus $h^i$ for various host materials, we find a common (*universal*) straight line of slope equal to unity. This equivalently means that the defect entropies in solids scale with the defect enthalpies irrespective of the process considered.

**Table 1**

Experimental values of the defect entropies (*s*) and enthalpies (*h*) in SrF$_2$. *F* denotes the bulk property discussed in the text.

| Process | Experimental | | | F | $s^i/F$ |
| --- | --- | --- | --- | --- | --- |
| | $h^i$ (eV) | $s^i$ (k-units) | $s^i/h^i$ ($10^{-4}$ K$^{-1}$) | ($10^{-4}$ K$^{-1}$) | (eV) |
| R$_1$ electrical relaxation: **Ce**[a] | 0.46 | 1.10 | 2.1 | 2.6 | 0.36 |
| R$_1$ electrical relaxation: **Eu**[a] | 0.457 | 1.18 | 2.3 | 2.6 | 0.39 |
| R$_1$ electrical relaxation: **Gd**[a] | 0.461 | 1.22 | 2.1 | 2.6 | 0.40 |
| anion-Frenkel formation | 2.05 | 6.35 | 2.7 | 2.6 | 2.1 |
| anion vacancy migration | 0.70 | 2.12 | 2.6 | 2.6 | 0.7 |
| anion interstitial migration | 0.93 | 4.34 | 4.0 | 2.6 | 1.4 |

[a]Measurements from Refs. [40, 41]

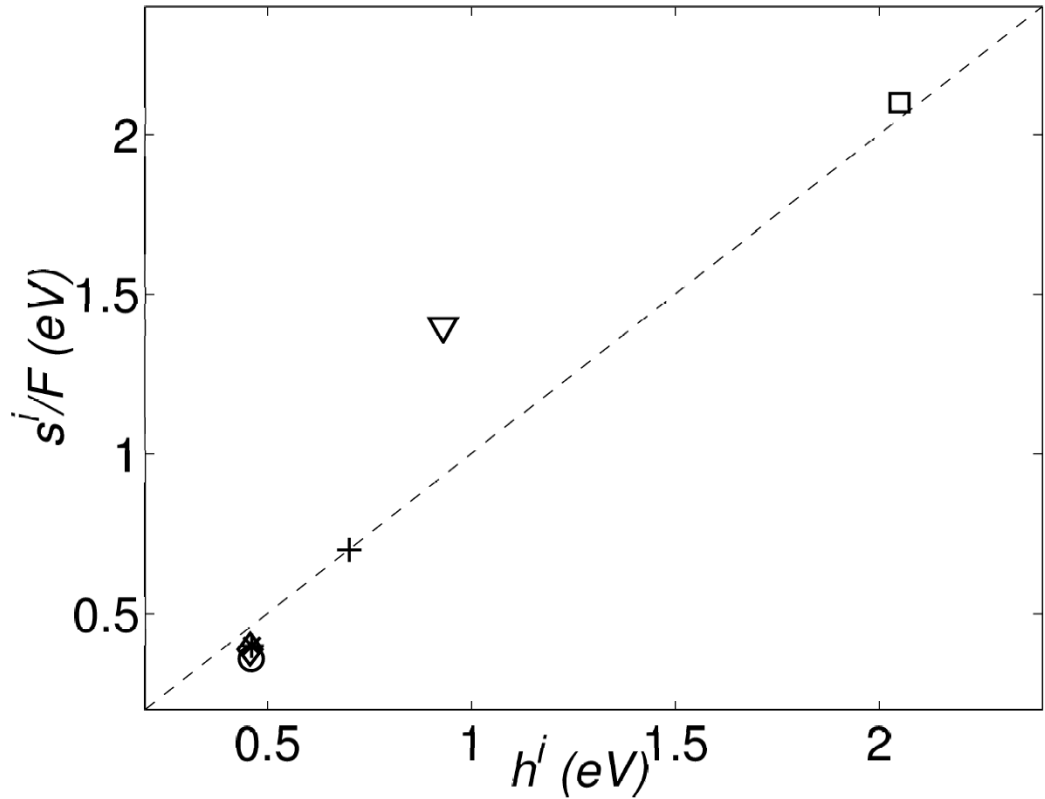

**Fig. 1**. The experimental values of $s^i/F$ versus the experimental defect enthalpies ($h^i$) for the following processes: $R_1$ electrical relaxation for $SrF_2$ doped with Ce (o), Eu (◊), Gd (*), anion-Frenkel formation (□), anion-vacancy migration (+), anion-interstitial migration (∇). The dashed straight line has slope equal to unity and is drawn for the sake of reader's convenience.